# Correction scheme for molecular total energies from quantum phase estimation under limited qubit resources

Nobuki Inoue [1,a] and Hisao Nakamura [1,a]

We propose a practical method for accurately evaluating molecular total energies using a hybrid approach that integrates fault-tolerant quantum computers with classical computing. Our scheme consists of two complementary components: quantum dominant orbital selection (QDOS) and subspace dynamical correlation (SDC). QDOS extracts only the essential active orbitals from the complete active space (CAS) configuration interaction (CI) state on a quantum computer, yielding a compact active space suitable for classical CASCI calculations. SDC then evaluates dynamical-correlation corrections for the CASCI energy using this compact state, which remains tractable on classical machines. To demonstrate that the CAS energy obtained on a quantum computer can be post-corrected by SDC, we examine two frameworks: multireference perturbation theory and tailored coupled-cluster theory. Our scheme enables effective treatment of relatively large molecular systems by combining limited quantum and classical resources.

## 1. INTRODUCTION

Ab initio quantum chemical calculations based on molecular orbital (MO) theory are standard tools for understanding molecular properties, functions, and fundamental chemical reactions. For quantitative simulations, evaluation of the electronic energy at the chemical accuracy level is often crucial. However, such calculations are computationally demanding because they require the incorporation of electron correlation effect, which is defined as the correction of the Hartree–Fock calculation; therefore, the application of these methods to complex molecular systems or chemical reaction processes remains limited. The configuration interaction (CI) method is the most fundamental technique for capturing electron correlation effects. In contrast to the Hartree–Fock or other mean-field methods, multi-configurational states, defined by the occupation of electrons in MOs and spin states, are used to construct CI wavefunctions and the resulting energy via variational principles. When all possible electronic configurations are adopted to expand the wavefunction, the method is referred to as full CI (FCI). [1,2] FCI provides a rigorous energy estimate based on the variational principle within a given basis set. However, the required computational resources increase exponentially with the size of the basis set (i.e., the atomic orbitals, equal to the number of total MOs), as well as the number of electrons in the system. Consequently, the application of FCI to practical chemical or material problems is nearly impossible, even when supercomputers are employed.

---

[1] *National Institute of Advanced Industrial Science and Technology (AIST), Tsukuba Central 2, 1-1-1 Umezono, Tsukuba, Ibaraki 305-8568, Japan*

[a] Authors to whom correspondence should be addressed: inoue.nbk@aist.go.jp and hs-nakamura@aist.go.jp

The exponential increase in the computational cost associated with FCI results from the exponential increase in the number of possible electronic configurations, such as quantum superpositions, as the number of molecular orbitals increases. Consequently, quantum computers, which explicitly manipulate quantum superposition, can exploit this exponentially expanding representational capacity, thereby offsetting the exponential growth in system size scaling. Therefore, the recent progress in quantum computers is expected to reduce the barriers to applying FCI calculations to practically relevant systems.

Currently, quantum computers are noisy intermediate-scale quantum (NISQ) devices [3] which directly use all qubits for computation without error correction. Several quantum chemistry algorithms have been proposed for the NISQ to establish the use of quantum computers. In the NISQ device algorithm, computational tasks are divided between those performed on quantum and conventional computers. Hereafter, we refer to conventional computers/computations as classical computers/computations to distinguish them from quantum computers. For example, the unitary coupled cluster (UCC) [4-7] method has emerged as a promising approach within the variational quantum eigensolver (VQE) framework [8]. Furthermore, quantum subspace expansion (QSE) [9] and quantum Krylov method [10-12] have been recently developed to improve the parameter optimization step in VQE. However, most NISQ algorithms for quantum chemical calculations encounter challenges, including the necessity for extensive number of "run and observation" cycles as the basis set size increases and errors introduced by quantum noise, which can affect the accuracy of the results. Therefore, the acceleration of quantum chemical calculations using the present NISQ-based algorithms for practical applications is an area of active debate.

In contrast to NISQ, fault-tolerant quantum computers (FTQCs) [13], which comprise qubits for computation and error correction, are immune to quantum noise. FTQCs are essentially universal computers, and quantum algorithms expected to have quantum advantages can be applied to them. For FCI energy calculations, adiabatic state preparation (ASP) [14,15] and quantum phase estimation (QPE) [16-18] algorithms are applicable with FTQC [19]. In addition, $n$ qubits can express multiconfigurational states comprising $2^n$ electronic configurations.

Therefore, classical computers are free from an exponential increase in computational resources. Based on these considerations, the development of quantum chemical algorithms integrating ASP and QPE has garnered renewed interest. Nevertheless, the realization of FTQC capable of performing FCI calculations with a large number of logical qubits remains a long-term objective, hindered by significant physical and engineering challenges.

An alternative approach to FCI is the use of multi-reference theory. The electron correlation effects in many-electron systems can be broadly classified into static and dynamical correlations. Static correlation is relevant when two or more electron configuration states are energetically quasi-degenerate, whereas dynamical correlation is related to electronic excitations to higher-energy MOs owing to electron-electron collisions. Energy correction owing to dynamical correlation is essential to provide accurate potential energy surface, particularly near the transition state. In multi-reference theory, the complete active space (CAS) configuration interaction (CASCI) is often adopted as the reference state (zeroth-order state). In the CASCI method, all MOs are classified as core, active, and virtual orbitals. The occupation numbers of the electrons in the core and virtual orbitals were fixed at 2 and 0, respectively. FCI is subsequently performed within the active orbitals. Therefore, CAS can be defined as the space spanned by all possible configurations of the active orbitals. The CASCI state includes the static correlation effect, and the dynamical correlation effect is incorporated by setting the CASCI wavefunction as the reference state. The dynamical correlation is incorporated by considering electronic configurations in the complementary space of the CAS, that is, excitation from the core to virtual electrons, via less computationally demanding calculations. Several methods based on multi-reference theory have been proposed such as the multi-reference configuration interaction method (MRCI) [20,21] multi-reference Møller–Plesset perturbation theory (MRMP) [22,23], complete active space perturbation theory (CASPT) [24,25], and *n*-electron valence state perturbation theory (NEVPT) [26-28]. However, in complex molecular systems, the number of active orbitals required to evaluate the static correlation increases rapidly owing to the increase in quasi-degenerate MOs, which makes it challenging to reduce the size of the CAS to calculate the CASCI wave function. Therefore, these multi-reference methods encounter difficulties similar to those of FCI as the molecular system size increases.

A promising approach involves a hybrid method using FTQC and classical computation employing multi-reference theory, where FTQC is used to calculate the CASCI energy, including static correlation, and the classical computation is performed to evaluate the dynamical correlation effect followed by the CASCI wavefunction. The problem with such a quantum-classical hybrid scheme is that the CASCI wave function (a more rigorous set of CI coefficients) is required to evaluate the dynamical correlation effect. However, the use of the CASCI wavefunction is not straightforward for the following two reasons: (i) many additional tasks are necessary to extract the data from the CASCI wavefunction, such as CI coefficients, which are encoded on qubits and (ii) the data must be stored on a classical computer to evaluate the correction terms of dynamical correlation. These data load (quantum) and storage (classical) operations may negate the expected quantum advantage. Although methods such as computational basis tomography (CBT) [29] have been proposed to estimate CI coefficients, including their phases from qubits, they require a substantially large number of measurements.

In this study, we developed a method based on the multi-reference theory in quantum-classical hybrid computation. We adopted an approach that approximates the CASCI wavefunction, which is *reconstructed* using a classical computer, and subsequently calculates the dynamical correlation. Specifically, we selected subsets of active orbitals to form a compact active space and constructed a new CAS based on this subspace. This compact active space is the subspace of the original CAS treated on a quantum computer. To distinguish the two types of CAS, we refer to the original CAS the reference CAS (rCAS) and CAS on the compact active space as the subspace CAS (sCAS). Furthermore, we refer to the CASCI state calculated using rCAS and sCAS as the rCASCI and sCASCI states, respectively. The sCASCI state is required to approximate the rCASCI state as accurately as possible. To minimize computational complexity, sCAS must be as small as possible and comprise dominant electron configurations derived from the dominant MOs in the rCASCI state.

In this study, we propose (i) a quantum-dominant orbital selection (QDOS) scheme to select the MOs for a compact set of active orbitals from rCASCI wavefunctions encoded on qubits and (ii) the subspace dynamical correlation (SDC) method, which evaluates the dynamical correlation energy of the rCASCI state obtained on a quantum computer by reconstructing and referencing the corresponding sCASCI wavefunction on classical computer. To distinguish our SDC method from conventional approaches, we use the terminology *"standard"* to identify the relating original dynamical correlation method, where the corrected state and the reference state are identical. In Section 2, we present the QDOS scheme and outline the SDC method. We adopted Møller–Plesset perturbation theory and tailored coupled cluster (TCC) methods to to incorporate the dynamical correlation energy into the rCASCI energy obtained via quantum computation. In Section 3, the numerical results and an analysis of the computational efficiency of the proposed scheme are discussed. Finally, Section 4 presents the summary.

## 2. Theoretical methods

### 2.1. Extract CASCI wavefunction on qubits by QDOS scheme

We assume that wavefunctions are encoded on qubits based on Jordan–Wigner mapping, and the MOs are fixed throughout the calculation. First, we set the core, active, and virtual orbitals to evaluate the rCASCI state, which was calculated and encoded using a quantum computer. For convenience, we labeled the core, active, and virtual orbitals of rCAS as core.(rCAS), act.(rCAS), and virt.(rCAS), respectively. The difference in the electron configurations in rCAS lies is the electronic occupation of the (act.(rCAS); therefore, we need to encode only part of the electron configuration constructed by act.(rCAS) into qubits. The Hamiltonian of rCASCI is expressed as

$$\hat{H}^{\mathrm{rCAS}} = E^{\mathrm{core}} + \sum_{pq}^{\mathrm{act.(rCAS)}} h_{pq}^{\mathrm{eff.}} \hat{a}_p^\dagger \hat{a}_q + \frac{1}{2} \sum_{pqrs}^{\mathrm{act.(rCAS)}} \left[ pq | rs \right] \hat{a}_p^\dagger \hat{a}_r^\dagger \hat{a}_s \hat{a}_q, \tag{1}$$

where $\hat{a}_p^\dagger$ and $\hat{a}_p$ are the creation and annihilation operators of an electron in the $p$-th MO, respectively. The first term in Eq. (1), $E^{\mathrm{core}}$, is the core electron energy, and is represented as

$$E^{\mathrm{core}} = \sum_{i}^{\mathrm{core(rCAS)}} h_{ii} + \frac{1}{2} \sum_{ij}^{\mathrm{core(rCAS)}} \left( \left[ ii | jj \right] - \left[ ij | ji \right] \right), \tag{2}$$

where $h_{pq}^{\mathrm{eff.}}$ is the effective one-electron integral expressed as,

$$h_{pq}^{\mathrm{eff.}} = h_{pq} + \sum_{i}^{\mathrm{core(rCAS)}} \left( \left[ pq | ii \right] - \left[ pi | iq \right] \right). \tag{3}$$

Here, $h_{pq}$ and $\left[ pq | rs \right]$ are the one- and two-electron integrals, respectively. In this study, we focused on evaluating the ground state for simplicity, and the rCASCI state was assumed to result from ASP and QPE. Here, we denote the numbers of active orbitals and active electrons for the general CAS as $N_{\text{act.(CAS)}}$ and $n_{\text{ele.(act.(CAS))}}$, respectively. The computational size of CASCI is characterized by both $N_{\text{act.(CAS)}}$ and $n_{\text{ele.(act.(CAS))}}$; hence, the notation CAS($N_{\text{act.(CAS)}}$, $N_{\text{act.(CAS)}}$) *etc.* is often used to identify the size of CAS.

In the QDOS scheme, the number of active orbitals and electrons in the sCAS are predetermined. The active orbitals for sCAS were selected based on measurements of the rCASCI state on a quantum computer. To obtain a good approximation of rCAS and reduce the number of active orbitals as much as possible, it is preferable to select MOs in act.(rCAS), whose electron occupancy is close to 1 via a spatial orbital. If the number of electron occupations is close to zero (2), such MOs can be relabeled to the virtual (core) orbital. Compared with extracting CI coefficients or analyzing the wave function data on qubits, estimating the occupancy of each MO orbital is significantly straightforward as it can be directly determined by measuring the electronic configuration of the rCASCI state on a quantum computer. Suppose that $N_j$ is the number of times the $j$-th digit is 1 in the $N_{\text{shot}}$ times the measurement of the quantum state after QPE. Given that each MO has $\alpha$ and $\beta$ spin labels, the estimated number of occupancies in $k$-th MO (more strictly, $k$-th active orbitals) $\eta_k$ is given by

$$\eta_k = \left(N_{2k-1} + N_{2k}\right)/N_{\text{shot}} . \tag{4}$$

Notably, the measurement of a qubit is a Bernoulli trial, and the standard error is

$$\sqrt{\frac{\bar{x}\left(1-\bar{x}\right)}{N_{\text{shot}}}} , \tag{5}$$

where $\bar{x}$ is mean value obtained by $N_{\text{shot}}$ times measurement. Therefore, the ratio of the standard error to the mean is extremely large when $\bar{x}$ is close to zero and that to $1-\bar{x}$ becomes substantially large when $\bar{x}$ is close to 1. Hence, assigning the occupation number of MOs by reasonable number of measurements is not straightforward when $\eta_k$ is nearly 0 or 2 because of statistical error. To address this statistical problem, we introduce the following criteria to evaluate $\eta_k$:

$$2-\frac{2\lambda}{\sqrt{N_{\text{shot}}}}<\eta_k\leq 2 \;\rightarrow\; \eta_k=2, \tag{6}$$

$$0\leq\eta_k<\frac{2\lambda}{\sqrt{N_{\text{shot}}}} \;\rightarrow\; \eta_k=0, \tag{7}$$

where $\lambda$ is an ad hoc parameter used to arrange the confidence interval, and we set it to $10^{-1/2}$ in this study.

The QDOS scheme can be summarized as follows.

i) Set the number of active orbitals ($N_{\text{act.(sCAS)}}$) and electrons ($n_{\text{ele.(act.(sCAS))}}$) of sCAS.

ii) Measure the rCASCI wave function encoded on a quantum computer for $N_{\text{shot}}$ times.

iii) Calculate $\eta_k$ given in Eq. (4), using the set of measured digits on each qubit obtained in step (ii).

iv) If the $\eta_k$ obtained in step (iii) is close to 0 or 2, update the $\eta_k$ using Eq. (6), (7).

v) Remove the $N_{\text{act.(rCAS)}}-N_{\text{act.(sCAS)}}-\left(n_{\text{ele.(act.(rCAS))}}-n_{\text{ele.(act.(sCAS))}}\right)/2$ orbitals with the largest values of $\eta_k$ from act(rCAS). If two orbitals have the same $\eta_k$ values, the orbital with the higher orbital energy must be removed in preference. These removed orbitals at the step (v) are now relabeled as virtual orbitals.

vi) Remove the $\left(n_{\text{ele.(act.(rCAS))}}-n_{\text{ele.(act.(sCAS))}}\right)/2$ orbitals with the smallest values of $\eta_k$ from act(rCAS). If two orbitals have the same $\eta_k$ values, the orbital with the lower orbital energy must be removed in preference. At step (vi) the removed orbitals are relabeled as core orbitals while the remaining orbitals constitute the active space of the sCAS.

Given that the number of configurations in sCAS is substantially smaller than that in rCAS, sCASCI can be easily calculated on a classical computer, that is, rCASCI is (approximately) reconstructed as sCASCI employing classical computation.

As a related approach to compress the reference space, the quantum selected CI (QSCI) method [30] was recently proposed. Here, we briefly note the differences between the QSCI scheme and our approach. In QSCI, wavefunctions encoded as qubits are repeatedly measured $N_{\text{shot}}$ times, and the configuration is determined to adopt in *selected* space (SS) owing to the frequency of each measured configuration being proportional to the square of the corresponding CI coefficient. However, "the members" of the configurations that were measured fewer times may not be consistently selected. In other words, the results of the $N_{\text{shot}}$ time measurements may vary, and even the number of configurations selected by the $N_{\text{shot}}$ time measurements can change. These issues become more prominent when different geometries are considered on a potential surface. This is because, depending on the geometry of the molecule, the number of configurations selected, that is, the quality of the SS, may change. Conversely, this problem of instability in the quality of a selected subspace can be resolved by predefining the size of the space.

Finally, although we introduced QDOS in this subsection to ensure the consistency between the rCASCI realized on an FTQC and sCASCI solutions, if the solution obtained on the FTQC can be reliably identified, purely classical active space selection schemes [31,32] can instead be employed to construct the sCAS. This would be helpful in avoiding the need for a large number of quantum measurements.

### 2.2. Subspace dynamical electron correlation method

In Section 2.1, we introduced the QDOS scheme to extract a compact set of active orbitals from the rCASCI wave function realized on a quantum computer, followed by the reconstruction of sCASCI as an approximation of CASCI on a classical computer. In this subsection, we present a method for evaluating the correction term of the dynamical correlation to the rCASCI state using the sCASCI wavefunction.

The rCASCI wavefunction on qubits as

$$\left|\Psi^{(\mathrm{rCAS})}\right\rangle = \sum_{A\in\mathrm{rCAS}} C_A^{(\mathrm{rCAS})}\left|A\right\rangle, \tag{8}$$

where $\left|A\right\rangle$ represents the Slater determinants included in the rCAS, and each of them corresponds to binary digits on quantum computers. The state of the sCASCI is expressed as follows:

$$\left|\Psi^{(\mathrm{sCAS})}\right\rangle = \sum_{A\in\mathrm{sCAS}} C_A^{(\mathrm{sCAS})}\left|A\right\rangle. \tag{9}$$

Although available wavefunction data correspond to the sCAS, the energy must be corrected to that of the rCASCI, rather than the sCASCI state. For example, when standard dynamical correlation methods are directly applied to the sCAS reference wavefunction, the resulting state already contains electronic configurations that belong to the rCASCI wavefunction. In other words, part of the dynamical correlation associated with the sCASCI reference state is already included in the QPE-evaluated energy as statical correlation. To prevent such double counting, it is sufficient that the rCAS–sCAS space is included in the reference space when accounting for dynamical electron correlation. In other words, we can achieve this by extending the size of CI coefficients referenced in the incorporation of dynamical electron correlation from sCAS to rCAS. By perturbatively supplementing the elements in the rCAS–sCAS space, we can obtain approximated rCASCI coefficient $\tilde{\mathbf{C}}^{(\mathrm{rCAS})}$ $\left(\left(\tilde{\mathbf{C}}^{(\mathrm{rCAS})}\right)_A = \tilde{C}_A^{(\mathrm{rCAS})}\right)$ as follows:

$$\tilde{C}_A^{(\mathrm{rCAS})} \sim C_A^{\mathrm{app.}} \Big/ \left\|\mathbf{C}^{\mathrm{app.}}\right\|, \tag{10}$$

$$C_A^{\mathrm{app.}} = \begin{cases} \sigma_A C_A^{(\mathrm{sCAS})} & \left(A\in\mathrm{sCAS}\right) \\ \sum_{B\in\mathrm{sCAS}} \left\langle A\middle|\hat{V}\middle|B\right\rangle \sigma_B C_B^{(\mathrm{sCAS})} f\left(E_{\mathrm{sCAS}}^{(0)} - E_A^{(0)}\right) & \left(A\notin\mathrm{sCAS}\right) \end{cases}, \tag{11}$$

with $\left(\left(\mathbf{C}^{\mathrm{app.}}\right)_A = C_A^{\mathrm{app.}}\right)$. Here, to capture the residual static electron correlation effects remaining in the rCAS–sCAS space, the perturbative energy denominator is replaced by a renormalization function,

$$f\left(\Delta\right) = \left(1-\exp\left(-s\Delta^2\right)\right)\Big/\Delta, \tag{12}$$

with the similar form of driven similarity renormalization group (DSRG) [33-35]. If the ordering of the MOs differs between the rCASCI and sCASCI calculations, sign factors $\sigma_A\left(=\pm 1\right)$ arising from the reordering of orbitals included in each Slater determinants are required. The perturbation is $\hat{V}=\hat{H}-\hat{H}^{(0)}$ and the zeroth energies are

$$E_{\mathrm{sCAS}}^{(0)}=\sum_{B\in \mathrm{sCAS}}\left\langle B\left|\hat{H}^{(0)}\right|B\right\rangle C_B^{(\mathrm{sCAS})*}C_B^{(\mathrm{sCAS})}, \tag{13}$$

and

$$E_A^{(0)}=\left\langle A\left|\hat{H}^{(0)}\right|A\right\rangle. \tag{14}$$

Using the approximated rCASCI coefficient $\tilde{\mathbf{C}}^{(\mathrm{rCAS})}$ on a classical computer, we can formulate the corrected total energy as

$$E_{\mathrm{TOT}}=E_{\mathrm{CASCI}}\left(\left|\Psi^{(\mathrm{CAS})}\right\rangle\right)+\Delta E_{\mathrm{corr.}}\left(\tilde{\mathbf{C}}^{(\mathrm{rCAS})}\right), \tag{15}$$

where the first and second terms on the right-hand side are the rCASCI total energy computed by the QPE and the electron correlation from the excitation configurations involving the core(rCAS) or virt.(rCAS), respectively. Notably, in the present scheme, an explicit separation between static and dynamical correlation terms is required. Consequently, the MRCI method is not suitable for direct application. By contrast, our proposed approach remains applicable even to formalisms that are not strictly multi-reference theory. In the following subsections, we examine two methods: MRMP and the tailored coupled cluster (TCC).

### 2.2.1. Application to MRMP method

MRMP formalism is a method of multi-reference theory in which the CASCI state is taken as the reference state. Within the second order Møller–Plesset perturbation, the second term on the RHS of Eq. (15) is straightforwardly given by

$$\Delta E_{\mathrm{MRMP2}}\left(\tilde{\mathbf{C}}^{(\mathrm{rCAS})}\right)=\sum_{A,B\in \mathrm{rCAS}}\sum_{I\notin \mathrm{rCAS}}\frac{\left\langle A\left|\hat{V}\right|I\right\rangle\left\langle I\left|\hat{V}\right|B\right\rangle}{E_{\mathrm{rCAS}}^{(0)}-E_I^{(0)}}\tilde{C}_A^{(\mathrm{rCAS})*}\tilde{C}_B^{(\mathrm{rCAS})}. \tag{16}$$

The zeroth energies for the denominator are

$$E_{\mathrm{rCAS}}^{(0)} = \sum_{A\in\mathrm{rCAS}} \left\langle A \middle| \hat{H}^{(0)} \middle| A \right\rangle \tilde{C}_A^{(\mathrm{rCAS})*} \tilde{C}_A^{(\mathrm{rCAS})}, \tag{17}$$

and

$$E_I^{(0)} = \left\langle I \middle| \hat{H}^{(0)} \middle| I \right\rangle, \tag{18}$$

respectively. Note that, due to the approximation in Eq. (11), the summations over the rCAS space in Eqs. (16) and (17) are limited to at most double excitations in the rCAS–sCAS space, making their evaluation tractable on classical computers. This method is referred to as Eqs. (16)–(18) proposed here "subspace-MRMP2."

### 2.2.2. Application to TCC method

The second example is the TCC. TCC is not strictly a multi-reference theory; however, it allows us to calculate the correction of the rCASCI energy as follows: When the cluster operator $\hat{T}$ in the TCC formulation is truncated by the single and double excitation terms (TCCSD), the second term on the RHS of Eq. (15) is expressed as

$$\begin{aligned}\Delta E_{\mathrm{TCCSD}}\left(\tilde{\mathbf{C}}^{(\mathrm{rCAS})}\right) &= \left\langle \Psi^{\mathrm{HF}} \middle| \hat{H} \exp\left(\hat{T}^{\mathrm{ext.}} + \hat{T}^{\mathrm{act.}}\left(\tilde{\mathbf{C}}^{(\mathrm{rCAS})}\right)\right) \middle| \Psi^{\mathrm{HF}} \right\rangle \\ &\quad - \left\langle \Psi^{\mathrm{HF}} \middle| \hat{H} \exp\left(\hat{T}^{\mathrm{act.}}\left(\tilde{\mathbf{C}}^{(\mathrm{rCAS})}\right)\right) \middle| \Psi^{\mathrm{HF}} \right\rangle,\end{aligned} \tag{19}$$

with

$$\hat{T}^{\mathrm{act.}}\left(\tilde{\mathbf{C}}^{(\mathrm{rCAS})}\right) = \hat{T}_1^{\mathrm{act.}}\left(\tilde{\mathbf{C}}^{(\mathrm{rCAS})}\right) + \hat{T}_2^{\mathrm{act.}}\left(\tilde{\mathbf{C}}^{(\mathrm{rCAS})}\right), \tag{20}$$

$$\hat{T}^{\mathrm{ext.}} = \hat{T}_1^{\mathrm{ext.}} + \hat{T}_2^{\mathrm{ext.}}, \tag{21}$$

where $\hat{T}^{\mathrm{act.}}$ is the cluster operator of the active part, which describes only the electron-excitation-related act.(rCAS), and $\hat{T}^{\mathrm{ext.}}$ are cluster operators of the external part that describe the electron excitation involving at least one orbital from the core(rCAS) or virt.(rCAS). Specifically, the operators in Eq. (20) are given by the rCASCI coefficients as follows:

$$\begin{aligned}\hat{T}_1^{\mathrm{act.}}\left(\tilde{\mathbf{C}}^{(\mathrm{rCAS})}\right) &= \sum_{S\in\mathrm{sCAS}} \left(\tilde{C}_S^{(\mathrm{rCAS})} \big/ \tilde{C}_0^{(\mathrm{rCAS})}\right) \hat{E}_S, \\ \hat{T}_2^{\mathrm{act.}}\left(\tilde{\mathbf{C}}^{(\mathrm{rCAS})}\right) &= \sum_{D\in\mathrm{sCAS}} \left(\tilde{C}_D^{(\mathrm{rCAS})} \big/ \tilde{C}_0^{(\mathrm{rCAS})}\right) \hat{E}_D - \frac{1}{2} \sum_{S,S'\in\mathrm{sCAS}} \left(\tilde{C}_S^{(\mathrm{rCAS})} \tilde{C}_{S'}^{(\mathrm{rCAS})} \Big/ \left(\tilde{C}_0^{(\mathrm{rCAS})}\right)^2\right) \hat{E}_S \hat{E}_{S'},\end{aligned} \tag{22}$$

where $\hat{E}_S\left(\hat{E}_{S'}\right)$ and $\hat{E}_D$ are the single and double excitation operators that generate excited configurations in the rCAS, respectively. The sets of coefficients $\tilde{C}_0^{(\mathrm{rCAS})}$, $\tilde{C}_S^{(\mathrm{rCAS})}\left(\tilde{C}_{S'}^{(\mathrm{rCAS})}\right)$, and $\tilde{C}_D^{(\mathrm{rCAS})}$ represent the rCASCI coefficients of the ground, single excitation, and double excitation configurations in rCAS, respectively. Furthermore, the operators in Eq. (21) are given as follows:

$$\hat{T}_1^{\mathrm{ext.}} = \sum_{S \notin \mathrm{rCAS}} T_S \hat{E}_S, \qquad \hat{T}_2^{\mathrm{ext.}} = \sum_{D \notin \mathrm{rCAS}} T_D \hat{E}_D, \tag{23}$$

where $\hat{E}_S$ and $\hat{E}_D$ are single- and double-excitation operators that generate excited configurations *not* included in rCAS, respectively. Amplitudes $T_S$ and $T_D$ were obtained using a standard CCSD scheme [36]. Here, we note that the expression (19) is similar to the correction energy term used in the corrected TCC given in Eq. (19) in Ref. [37] rather than the original TCC [38]. Hereinafter, we refer to the TCCSD scheme based on Eqs. (19)–(22) "subspace-TCCSD."

The subspace-TCCSD can be extended to the TCCSD(T) framework, which incorporates the three-electron excitation term (T) by perturbation, by rewriting Eq. (15) as

$$\Delta E_{\mathrm{TCCSD(T)}}\left(\tilde{\mathbf{C}}^{(\mathrm{rCAS})}\right) = \Delta E_{\mathrm{TCCSD}}\left(\tilde{\mathbf{C}}^{(\mathrm{rCAS})}\right) + \Delta E_{(\mathrm{T})}\left(\hat{T}^{\mathrm{ext.}}\right). \tag{24}$$

The (T) correction can be obtained by excluding $\hat{T}^{\mathrm{act.}}$ from the total amplitude to avoid double counting of triple excitations in rCAS, that is, by applying only $\hat{T}^{\mathrm{ext.}}$ to the perturbation term in the standard CCSD(T) [39,40]. The perturbative corrections using Eq. (24) can be denoted as "subspace-TCCSD(T)."

## 3. Numerical results

In this section, we analyze the numerical stability of our QDOS method by comparing it with QSCI, and evaluate the accuracy of the subspace-MRMP/TCC method against *standard* MRMP/TCC methods by calculating the potential energy curves. We consider the $F_2$ and $N_2$ molecules using the cc-pVDZ basis set, which is often employed to assess the accuracy of proposed quantum chemical methods. In addition, we checked the error in the proposed method by comparing the results with the exact energy, specifically the FCI energy, for the ethylene molecule using the STO-3G basis set.

Since QPE calculations are not feasibly using present quantum computers, we prepared the quantum state related to the rCASCI, which was obtained using classical computers. The QDOS protocol was then implemented to extract the necessary quantum data utilizing the quantum simulator Qulacs [41]. All quantum chemical calculations were performed employing our in-house program, which was adapted from the PySCF [42] program.

### 3.1. Numerical stability of QDOS scheme

First, we focus on quality of the subspace extracted from the rCASCI wavefunction on quantum computer, i.e., the quantum data obtained via QDOS. We present the calculated potential energy curves of the $F_2$ molecule using subspace-MRMP2. We examined the CI coefficients to evaluate the dynamical correlation obtained by the QDOS and QSCI schemes. The protocol is summarized as follows. First, we performed CASCI calculations using the Hartree–Fock (HF) orbital by setting rCAS as CAS(10e,11o) and embedding the CASCI wave function in qubits as quantum data on a quantum simulator. These steps correspond to quantum-state (rCASCI) preparation for QDOS. The states of each qubit on the quantum simulator were sampled based on the QDOS scheme to determine the sCAS for subsequent SDC calculations. We also used the SS to apply the SDC method to the QSCI scheme. The number of measurements, $N_{\text{shot}}$, was set to 1000 for both QDOS and QSCI. sCAS was defined as CAS(2e,2o). When constructing approximate CI coefficients of rCAS for the SDC method from the CI coefficients obtained in the sCAS/SS space based on Eqs. (10) and (11), the flow parameter $s$ in the renormalization function [Eq. (12)] was set to $s = 0.0$ for QSCI, and to both $s = 0.0$ and 0.5 for QDOS. Note that $s = 0.5$ is the standard parameter adopted in the DSRG calculations[35], whereas $s = 0.0$ corresponds to the case in which all perturbatively approximated terms in Eq. (11) are set to zero. When it is necessary to explicitly specify the value of $s$, we use the notation such as "SDC($s$ = *)" hereafter. To verify numerical stability, the above procedure was repeated 100 times for each QDOS and QSCI value. The results are shown in Fig. 1. The solid red curves show the mean values of the subspace-MRMP2 energies, whereas the gray area shows the range of the obtained energies. In QSCI+SDC($s$ = 0.0), the variability of calculated energies is typically approximately 0.01 a.u. as shown in Fig. 1 (a). This discrepancy from the mean value was observed when the bond length was greater than the equilibrium position. In other words, a single calculation of the subspace-MRMP2 using the QSCI+SDC($s$ = 0.0) scheme introduces a non-negligible error in evaluating the activation barrier or dissociation energy. By contrast, the QDOS+SDC($s$ = 0.0) results in Fig. 1 (b) indicate that the fluctuation in the evaluated energies is substantially smaller than that of QSCI (~ 0.0013 a.u.). Notably, the adopted CAS(2e,2o) is a minimal multiconfigurational space and poses a *tough condition* for the QDOS. In such a small sCAS, it is imperative for the QDOS scheme to precisely identify most important orbitals as active, as tolerance for inaccuracies is minimal in the $\eta_k$ estimation. Here, we note that QDOS+SDC scheme with nonzero $s$ can perturbatively correct deviations in the measurement results. In fact, as shown in Fig. 1 (c), for QDOS+SDC($s$ = 0.5), the fluctuation in the evaluated energies remains at approximately 0.0007 a.u., which is about half of that observed for QDOS+SDC($s$ = 0.0). Therefore, QDOS+SDC is numerically stable, and the stability is further enhanced for nonzero values of $s$; in other words, it is expected to be robust with respect to fluctuations in the measurement results.

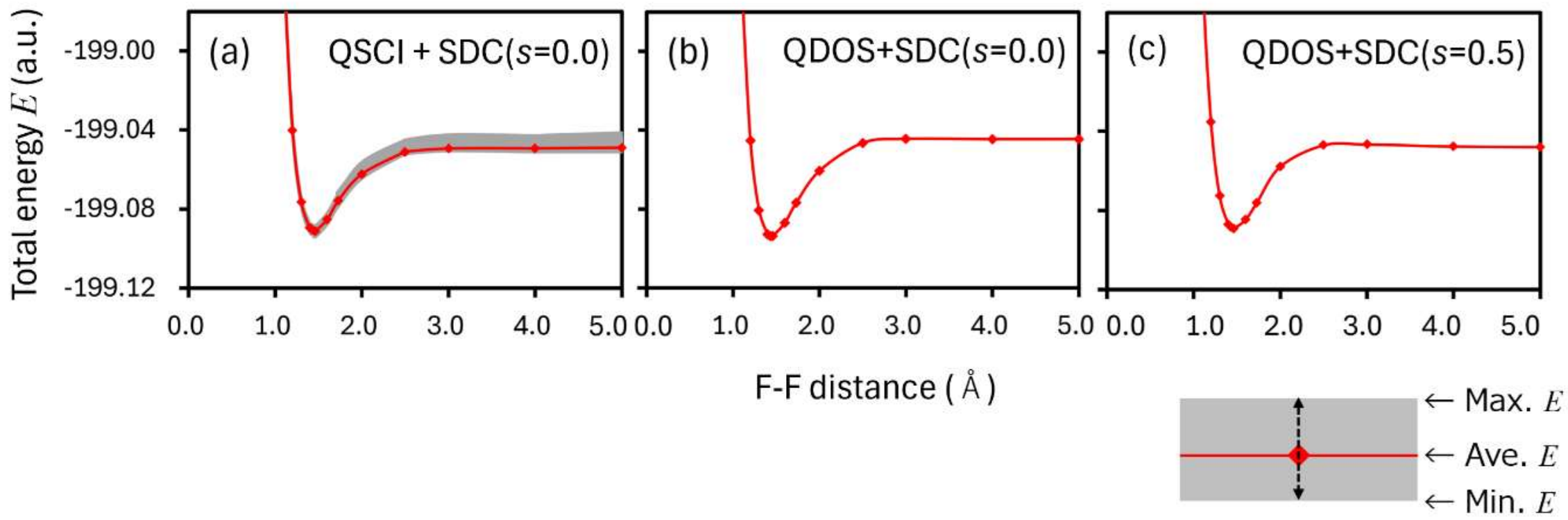

**Fig. 1.** Comparison of ground state potential energy curves of $F_2$ molecule calculated via subspace-MRMP2 method using sCAS/SS selected by QDOS/QSCI.

## 3.2. Validity of subspace dynamical correlation (SDC) methods

As stated in Section 2, the SDC methods evaluate the correction of rCAS energy using the sCASCI wavefunction rather than rCASCI. In standard dynamical correlation methods, such as MRMP and TCC, the correction of rCAS energy is evaluated using the rCASCI

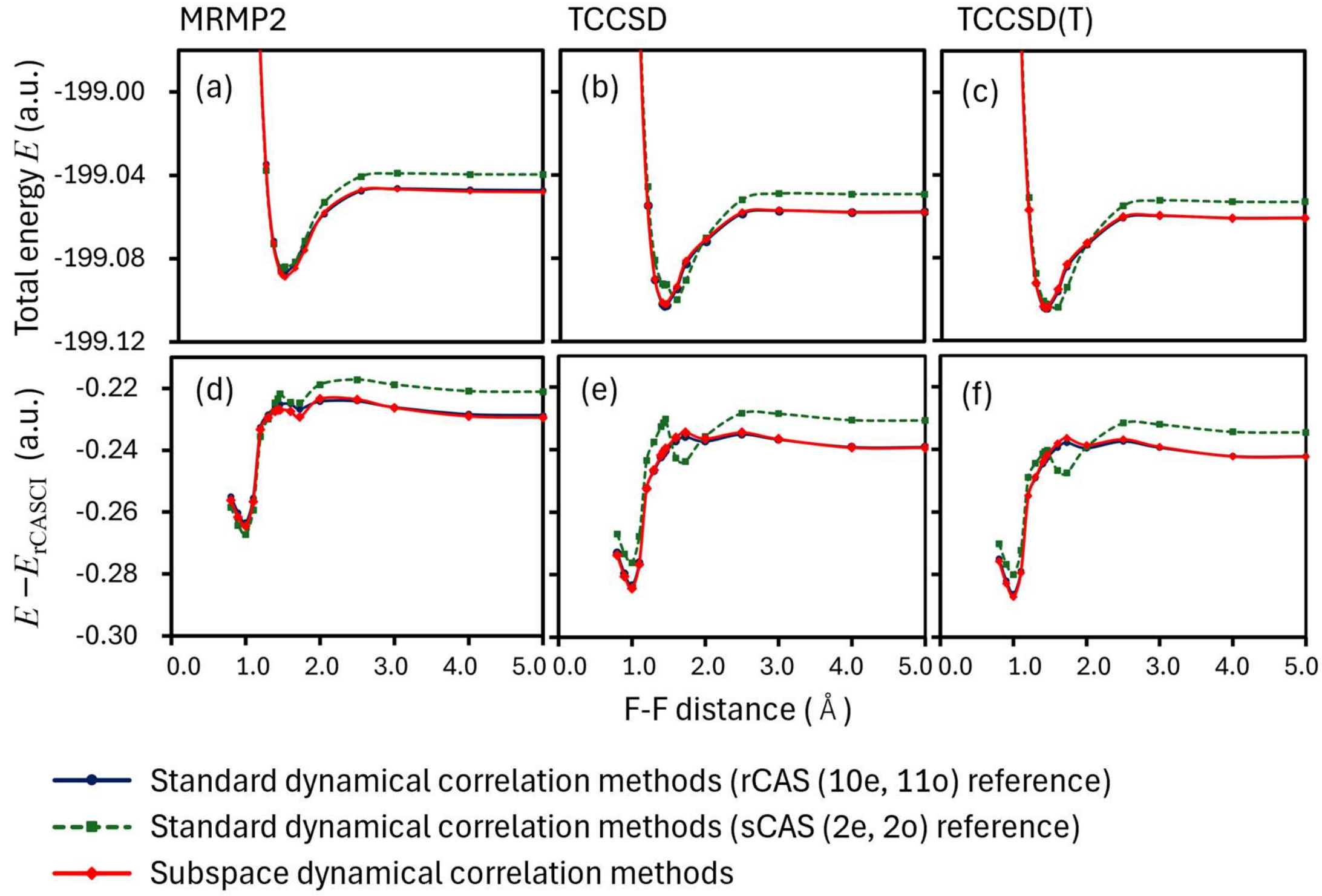

**Fig. 2.** Comparison of ground state potential energy curves of $F_2$ molecule calculated via standard and SDC methods.

wavefunction.

Hence, it is necessary to analyzing the accuracy of the dynamical correlation energy to rCAS energy computed employing SDC methods (subspace-MRMP/TCC) and compare it to that of the standard dynamical correlation methods (MRMP/TCC).

In addition to the $F_2$ molecule, we calculated the potential energy curves for the $N_2$ molecules, where the rCAS and sCAS of $N_2$ were CAS(10e,12o) and CAS(6e,6o), respectively. The value of $N_{\mathrm{shot}}$ was set to 1000 and $s = 0.5$ is adopted for the flow parameter.

The results of $F_2$ are summarized in Fig. 2. In Fig. 2 (a), we plotted the potential energy curves of $F_2$ using rCASCI-referenced standard MRMP2 (solid navy blue curve), sCASCI-referenced standard MRMP2 (green dashed curve) and subspace-MRMP2 (solid red curve). While the curves of the rCASCI-referenced standard MRMP2 and subspace-MRMP2 are almost indistinguishable, a clear discrepancy is observed between the rCASCI-referenced and sCASCI-referenced standard MRMP2 curves. Specifically, the energy differences relative to the rCASCI-referenced standard MRMP2 are −0.0018 a.u. at 1.46 Å and −0.00077 a.u. at 5.0

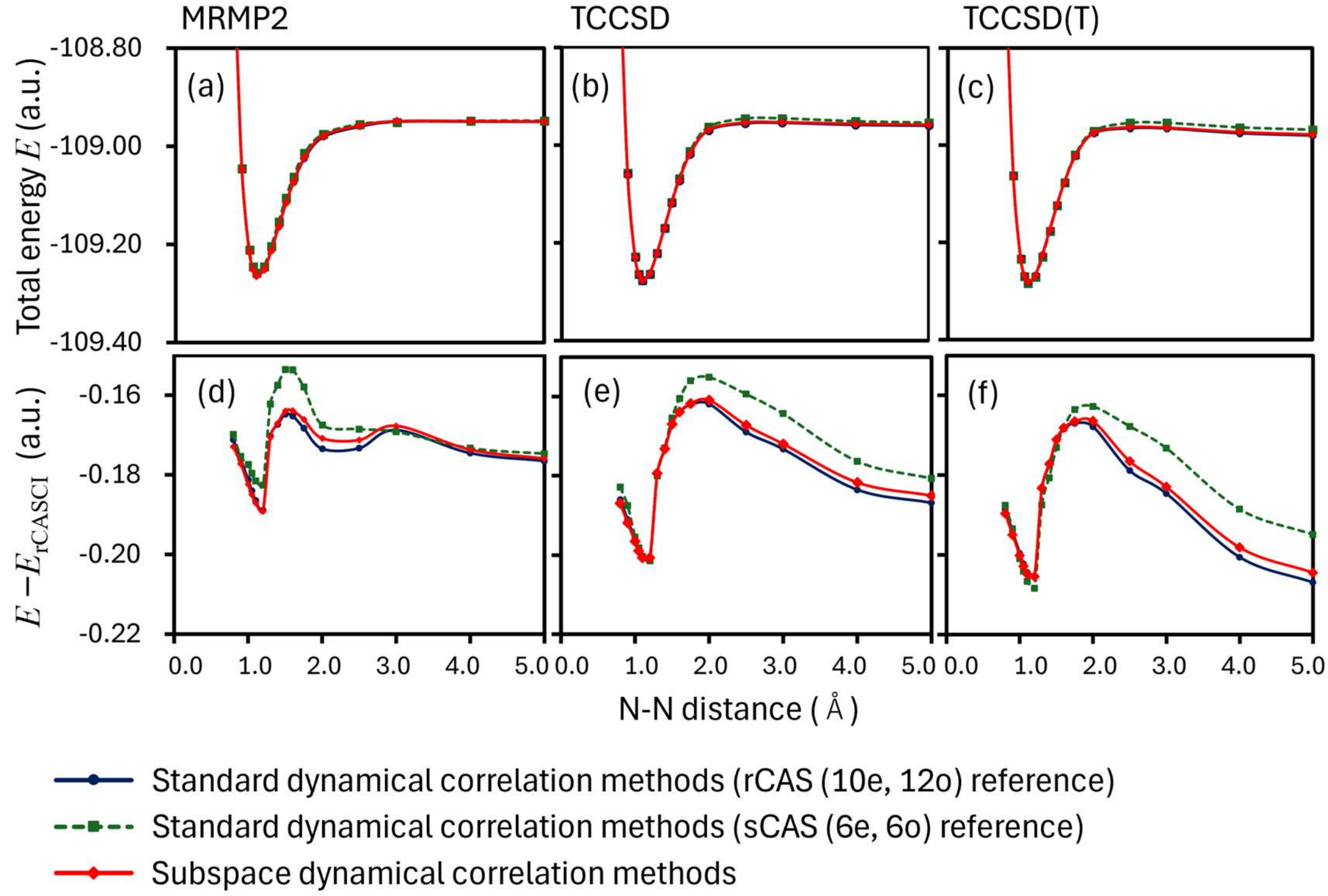

**Fig. 3.** Comparison of ground state potential energy curves of $N_2$ molecule calculated via standard and SDC methods.

**Table 1.** Total energy difference from FCI total energy of ethylene. The reference FCI total energy is −77.2351 (a.u.). The sizes of the spaces for sCAS, rCAS, and FCI is (2e,2o), (4e,4o), and (16e,14o).

| Method | $\Delta E$ (a.u.) |
|---|---|
| rCASCI | 0.1159 |
| MRMP2 (rCAS ref.) | 0.0285 |
| MRMP2 (sCAS ref.) | 0.0298 |
| subspace-MRMP2 | 0.0288 |
| TCCSD (rCAS ref.) | −0.0028 |
| TCCSD (sCAS ref.) | −0.0036 |
| subspace-TCCSD | −0.0026 |
| TCCSD(T) (rCAS ref.) | −0.0035 |
| TCCSD(T) (sCAS ref.) | −0.0042 |
| subspace-TCCSD(T) | −0.0034 |

Å for subspace-MRMP2, whereas they are 0.0032 a.u. at 1.46 Å and 0.0076 a.u. at 5.0 Å for the sCASCI-referenced standard MRMP2. These results indicate that subspace-MRMP2 reproduces the rCASCI-referenced standard MRMP2 with near chemical accuracy. Fig. 2 (d) shows the correction energies related to the rCASCI energy. The ratio of the dynamical correlation energy recovered by the subspace-MRMP2 method to that of the rCASCI-referenced standard MRMP2 method was 100.8 % at 1.46 Å and 100.3 % at 5.00 Å. This indicates that the subspace-MRMP2 method successfully incorporated the electron correlation effects from the core and virtual orbitals that could not be included in the qubits. Thus, the dynamical correlation to the rCASCI energy recovered by the rCASCI-referenced standard MRMP2 was largely preserved by the subspace-MRMP2. The potential curves for rCASCI- and sCASCI-referenced standard TCCSD and subspace-TCCSD are shown in Fig. 2 (b), and the correction energies to rCAS related to the above are plotted in Fig. 2 (e). The dynamical correlation energy by the rCASCI-referenced standard TCCSD aligns closely with that of the subspace-TCCSD, as in the case of the rCASCI-referenced standard MRMP2 and subspace-MRMP2. The energy difference 0.00086 a.u. at 1.46 Å and −0.00022 a.u. at 5.0 Å and the recovery rate of the dynamical correlation energy is 99.6 % at 1.46 Å and 100.1 % at 5.00 Å. We observed a similar reasonable agreement between the rCASCI-referenced TCCSD(T) and subspace-TCCSD (T) results shown in Fig. 2 (c) and (f).

Next, we consider the $N_2$ molecule, whose dissociation involves a triple bond and presents a significant challenge for accurate electronic structure modeling. In Fig 3, the calculated results for the $N_2$ molecule are shown using the same notation as in Fig. 2. For the $N_2$ molecule, at 1.1 Å, the dynamical correlation energies obtained by the SDC methods exhibit an accuracy that clearly exceeds chemical accuracy relative to the standard methods. In contrast, at 5.0 Å, although subspace-MRMP surpasses chemical accuracy (0.00059 a.u.), subspace-TCCSD and subspace-TCCSD(T) remain *high* accuracy (0.0018 and 0.0025 a.u., respectively). Therefore, the SDC methods demonstrate performance that clearly surpasses that of the sCASCI-referenced standard dynamical correlation methods. Consequently, the method demonstrated sufficient accuracy to serve as a viable alternative in situations where standard approaches are computationally infeasible.

Finally, we present the results of the comparison with the FCI energy by application to an ethylene molecule. Here, we adopted the STO-3G basis set owing to the limitations of computer resources, while the possible number of electronic configurations of ethylene is higher than those of $N_2$ and $F_2$ with the STO-3G basis set. Table 1 lists the deviations in the total energies calculated using each method from the FCI. Each SDC method significantly improves the rCASCI energy. The energy difference between SDC and rCASCI-referenced standard methods was within 0.0003 a.u.

Based on the above calculations, we conclude that the SDC method provides a reliable approximation of standard dynamical correlation method with CASCI reference state, while avoiding reliance on specific approaches such as MRMP or TCC. Notably, applying conventional MRMP/TCC to large rCAS spaces is challenging not only on classical computers but also on quantum computers, since the amount of quantum data (CI coefficients) that must be loaded grows rapidly with system size. Hence, subspace-MRMP/TCC are significantly promising as alternative approaches to families of CASCI referenced dynamical correlation methods.

## 4. Summary

We proposed a quantum chemical calculation method using quantum (FTQC)-classical hybrid computation, which comprises two key components: the QDOS scheme and SDC method. In the QDOS scheme, a relatively small CAS is selected as a subspace by effectively extracting the required data from the wavefunction encoded on a quantum computer, which is a large quantum state. The SDC method corrects the large CASCI energy using a quantum computer for postprocessing with reasonable computational resources. In this step, approximate CI coefficients in the subspace are reconstructed and referenced to evaluate the dynamical correlation effects on the large CAS. We adopted two standard dynamical correlation methods, MRMP and TCC, and derived concrete expressions for the SDC method.

Given that the QDOS scheme avoids extracting wavefunction data (CI coefficients) directly, it is robust to inconsistency of the selected subspace by quantum fluctuations or changes in molecular geometry and requires less effort to read data encoded on a quantum computer. We also demonstrated that energy correction by large-scale CASCI followed by the standard dynamical correlation method can be reproduced employing the SDC method. Furthermore, we emphasize that we do not require a large CASCI wavefunction, which is assumed to be embedded on a quantum computer. CASCI, followed by the standard dynamical correlation method, is presently one of the most reliable methods for calculating accurate potential energy surfaces, where FCI is not applicable to classical computers. Because the required number of qubits for CASCI is typically substantially smaller than that for FCI, we believe that our approach offers a practical use case for early FTQC, where only a limited number of logical qubits are available.

Finally, we discuss the problem of MOs playing the role of methods that rely on subspaces. Owing to the recalculation of CASCI by the active MOs selected by the QDOS, it was demonstrated that our proposed method yields significantly smoother potential energy surfaces compared with the other schemes over the entire region. However, this may not be expected, particularly when static correlation from many electronic configurations is relevant (i.e., many MOs relating to the ground state or the target excited state are quasi-degenerated). Most of the present quantum algorithms, such as the multi-configurational self-consistent field theory, do not provide significant advantages for the optimization of MOs. Hence, defining a consistent subspace constructed by a fixed small number of dominant electronic configurations (or dominant MOs) over the entire molecular conformation is not necessarily feasible, even when the MO transformation within the active MOs can be effectively performed using a quantum algorithm. In such cases, the size of the subspace must be increased to evaluate an accurate and smooth potential energy surface, and post-processing using classical computers may become impractical. Practical schemes, including the optimization of MOs using FTQC, will be required for subspace approaches based on the MO theory in future studies.

## AUTHOR DECLARATIONS

### Conflict of interest

The authors have no conflicts to disclose.